\documentclass[aps,prl,twocolumn,10pt,superscriptaddress,nofootinbib,nobibnotes,longbibliography]{revtex4-1}

\usepackage{amssymb}
\usepackage{graphicx}
\usepackage{amsmath}
\usepackage{mathrsfs}
\usepackage[colorlinks=true,linkcolor=blue,citecolor=blue,urlcolor=blue]{hyperref}
\usepackage{subfigure}
\usepackage{multirow}
\usepackage{setspace}
\usepackage{verbatim}
\usepackage{float}
\usepackage{color}
\usepackage{ulem}
\usepackage[utf8]{inputenc}
\usepackage[table,xcdraw]{xcolor}
\usepackage{makecell}
\usepackage{url}

\begin{document}

\title{Relic gravitational waves from primordial gravitational collapses}

\author{Xiang-Xi Zeng}
\email{zengxiangxi@sjtu.edu.cn (first two authors contribute equally)}
\affiliation{Institute of Theoretical Physics, Chinese Academy of Sciences (CAS), Beijing 100190, China}
\affiliation{School of Physical Sciences, University of Chinese Academy of Sciences (UCAS), Beijing 100049, China}
\affiliation{State Key Laboratory of Dark Matter Physics,
Tsung-Dao Lee Institute and School of Physics and Astronomy,
Shanghai Jiao Tong University, Shanghai 201210, China}
\affiliation{Key Laboratory for Particle Astrophysics and Cosmology (MOE),
and Shanghai Key Laboratory for Particle Physics and Cosmology,
Shanghai Jiao Tong University, Shanghai 201210, China}

\author{Zhuan Ning}
\email{ningzhuan17@mails.ucas.ac.cn (first two authors contribute equally)}
\affiliation{School of Fundamental Physics and Mathematical Sciences, Hangzhou Institute for Advanced Study (HIAS), University of Chinese Academy of Sciences (UCAS), Hangzhou 310024, China}
\affiliation{Institute of Theoretical Physics, Chinese Academy of Sciences (CAS), Beijing 100190, China}
\affiliation{University of Chinese Academy of Sciences (UCAS), Beijing 100049, China}

\author{Zi-Yan Yuwen}
\email{ziyan.yuwen@apctp.org}
\affiliation{Institute of Theoretical Physics, Chinese Academy of Sciences (CAS), Beijing 100190, China}
\affiliation{School of Physical Sciences, University of Chinese Academy of Sciences (UCAS), Beijing 100049, China}
\affiliation{Department of Physics, Stellenbosch University, Matieland 7602, South Africa}
\affiliation{Asia Pacific Center for Theoretical Physics (APCTP), Pohang 37673, Korea}

\author{Shao-Jiang Wang}
\email{schwang@itp.ac.cn (corresponding author)}
\affiliation{Institute of Theoretical Physics, Chinese Academy of Sciences (CAS), Beijing 100190, China}
\affiliation{Asia Pacific Center for Theoretical Physics (APCTP), Pohang 37673, Korea}

\author{Heling Deng}
\email{hd2586@columbia.edu}
\affiliation{Department of Astronomy, Columbia University, New York, NY 10027, USA}

\author{Rong-Gen Cai}
\email{caironggen@nbu.edu.cn}
\affiliation{Institute of Fundamental Physics and Quantum Technology, \& School of Physical Science and Technology, Ningbo University, Ningbo, 315211, China}

\begin{abstract}
A large primordial density perturbation of the Hubble scale will gravitationally collapse, generating an outgoing sound shell, whether or not a primordial black hole (PBH) is formed. In this Letter, we report a hybrid numerical analysis of the stochastic gravitational wave background induced by the collision of sound shells in the early Universe. The peak frequency and amplitude in the GW spectrum depend on the Hubble horizon and the abundance of sound shells. Abundant density perturbations would lead to GW backgrounds potentially detectable for future pulsar timing arrays and ground-based/space-borne detectors. For those perturbations that collapse into PBHs, future null detection of the corresponding high-frequency GW background could put new observational constraints on those PBHs that have already evaporated.
\end{abstract}
\maketitle

\textit{\textbf{Introduction.}---} 
Recently, various Pulsar Timing Array (PTA) collaborations (NANOGrav~\cite{NANOGrav:2023gor}, EPTA~\cite{EPTA:2023fyk}, PPTA~\cite{Reardon:2023gzh}, and CPTA~\cite{Xu:2023wog}) have announced possible detections of stochastic gravitational-wave backgrounds (SGWBs). With more upcoming GW detectors (such as space-borne detectors like LISA~\cite{LISA:2017pwj}, Taiji~\cite{Hu:2017mde, Ruan:2018tsw}, and TianQin~\cite{TianQin:2015yph}, as well as the more advanced DECIGO~\cite{Kawamura:2006up} and BBO~\cite{Corbin:2005ny}, along with ground-based detectors like Einstein Telescope (ET)~\cite{ET:2019dnz} and Cosmic Explorer (CE)~\cite{CE:2019iox}), we might embrace many SGWBs~\cite{Caprini:2018mtu} with rich new physics~\cite{Cai:2017cbj,Bian:2021ini}. Therefore, it is both observationally and theoretically appealing to accurately calculate and analyze the signal of SGWB.

It has long been expected~\cite{Hawke:2002rf,Musco:2004ak,Musco:2012au,Bloomfield:2015ila,Garcia-Bellido:2019vlf, Carr:2019hud} that an outgoing sound shell~\cite{Deng:2017uwc,Deng:2018cxb,Deng:2020pxo,Deng:2020mds} will form along with primordial black hole (PBH) formation from gravitational collapse of a sufficiently large density perturbation of horizon size. In this \textit{Letter}, we report a hybrid numerical analysis of the SGWB induced by the superposition of sound waves from gravitational collapses of large density perturbations of horizon size in the early Universe. This provides a crucial extension to studies of scalar-induced gravitational waves (SIGWs), as it captures, for the first time, the inherently non-perturbative dynamics of the source. The details of numerical simulations are presented separately in two accompanying long papers~\cite{Ning:2025ogq,Ning:2025yvj} on the long-time evolution of such large density perturbations. This \textit{Letter} will focus on the GW phenomenology even if these density perturbations are not large enough to form PBHs.

\textit{\textbf{Sound waves from density perturbations.}---}
PBHs can form in the early Universe if the perturbation amplitude exceeds a critical threshold~\cite{Carr:1974nx, Hawking:1971ei}. For those density perturbations of horizon size that are not sufficiently dense to result in PBH formation, bounces occur, and a shell of density contrast dubbed sound wave propagates outwards, as shown in previous numerical simulation results~\cite{Hawke:2002rf, Musco:2004ak, Bloomfield:2015ila, Musco:2012au}. In the accompanying long paper~\cite{Ning:2025ogq}, we find that the sound wave could also form for the case with PBH formation, and hence, we will follow the numerical treatment of the PBH formation.

\begin{figure*}
    \centering
    \includegraphics[width=\textwidth]{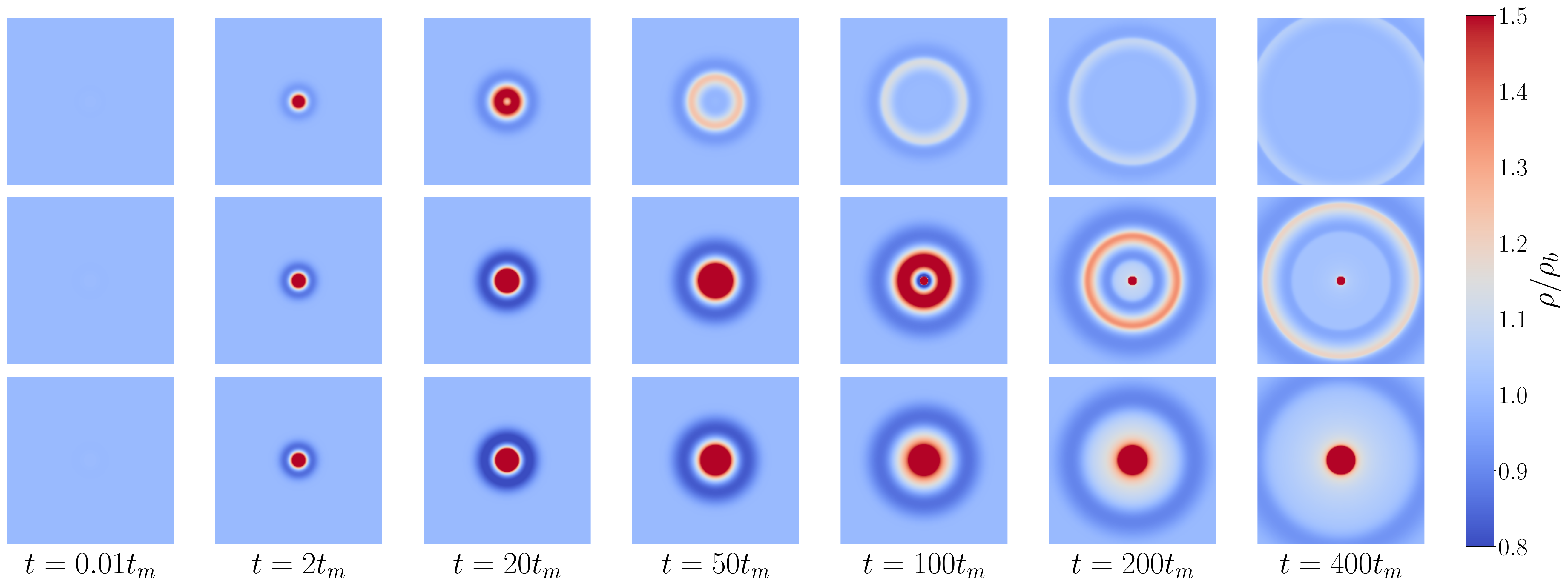}
    \caption{Evolution of the energy density in the process of PBH formation. The \textit{upper} panel describes the sub-critical case with $\mu=0.4$. In this case, only a sound wave is formed, as the perturbation is not dense enough. The \textit{middle} panel describes the near-critical case with $\mu=0.8$. Due to the large gradient of fluid pressure, both underdense and overdense structures appear. The \textit{lower} panel denotes the super-critical case with $\mu =0.9$. In this case, there is only an underdense structure.}
    \label{fig:simulation}
\end{figure*}

\begin{figure*}
    \centering
    \includegraphics[width=\textwidth]{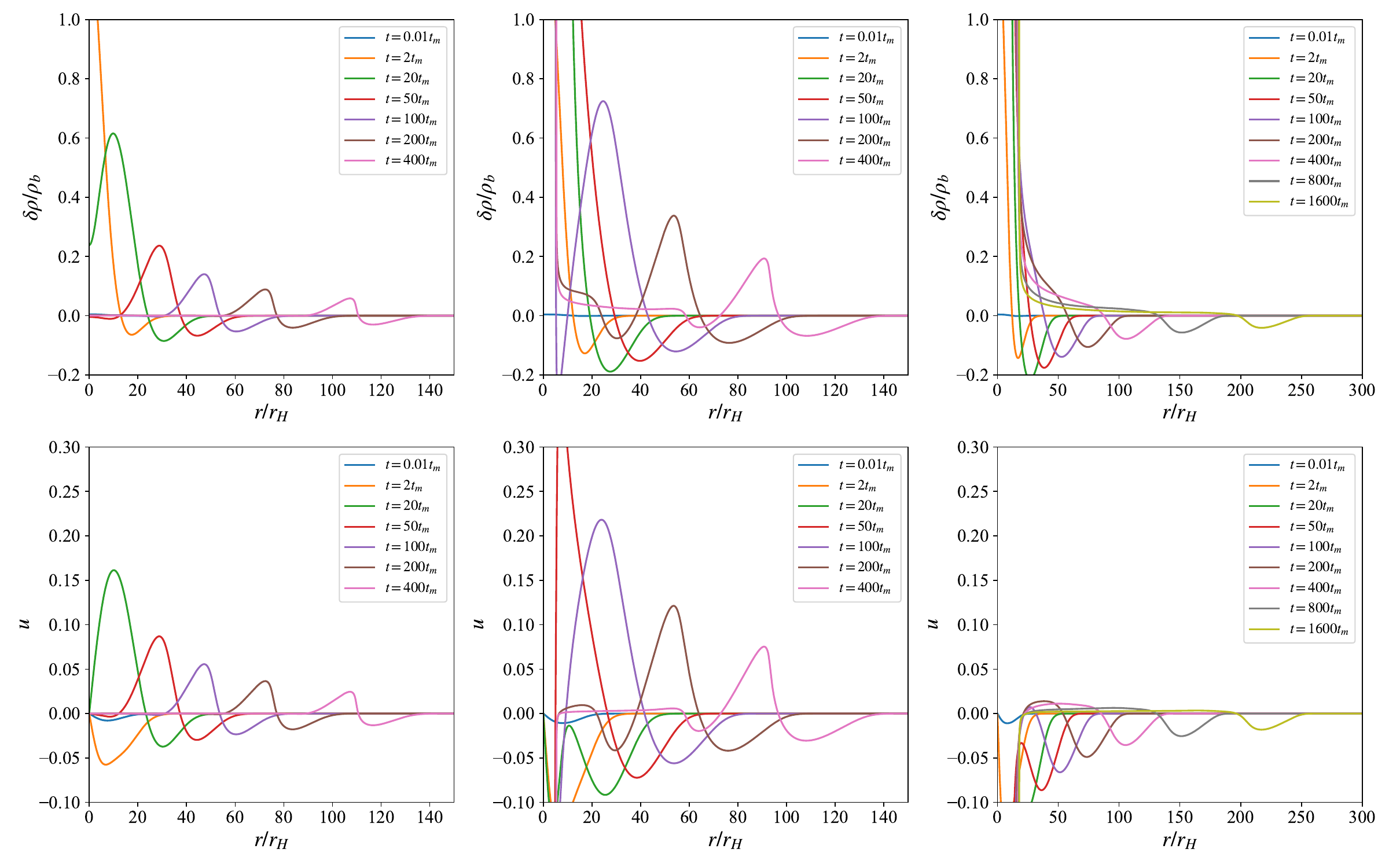}
    \caption{The density contrasts $\delta$ (top row) and the comoving radial velocity $u$ (bottom row) of the fluid for the sub-critical case with $\mu=0.4$ (left column), near-critical case with $\mu=0.8$ (middle column), and super-critical case with $\mu=0.9$ (right column), where the PBHs are formed around $t=44.75t_m$ and $t=19.83t_m$ with the initial PBH comoving radius $r_\mathrm{PBH} = 3.08r_H$ and $r_\mathrm{PBH} = 5.32r_H$, respectively.}
    \label{fig:profiles}
\end{figure*}

Assuming spherical symmetry, numerical simulation conventionally solves the Misner-Sharp equations~\cite{Misner:1964je} for the motion of the relativistic fluid with a perfect fluid ansatz for the energy-momentum tensor $T^{\mu\nu} = (p+\rho) u^\mu u^\nu + pg^{\mu\nu}$, where $u^{\mu}$ is the fluid four-velocity and $p$ and $\rho$ are related by an equation-of-state parameter $\omega\equiv p/\rho$. The line element with spherical symmetry reads $\mathrm{d}s^2 = -A(r, t)^2 \mathrm{d}t^2 + B(r, t)^2 \mathrm{d}r^2 + R(r, t)^2\mathrm{d}\Omega^2$ with $\mathrm{d}\Omega^2 = \mathrm{d}\theta^2 + \mathrm{sin}^2(\theta)\mathrm{d}\phi^2$. At superhorizon scales, the metric can be rewritten using the long-wavelength approximation~\cite{Salopek:1990jq, Polnarev:2006aa, Polnarev:2012bi, Harada:2015yda},
\begin{equation}
    \mathrm{d}s^2 = -\mathrm{d}t^2 + a(t)^2e^{2\zeta(r)}\left(\mathrm{d}r^2 + r^2\mathrm{d}\Omega^2\right),
\end{equation}
where $\zeta(r)$ encodes the information of the initial curvature perturbations that further induce density perturbations. In this \textit{Letter}, we simply follow the conventional treatment by adopting a Gaussian profile for the initial curvature perturbation,
\begin{align}
    \zeta(r) = \mu e^{-(r/r_m)^2},
\end{align}
where $r_m$ is the length scale of the perturbation and $\mu$ denotes its peak amplitude. The Gaussian profile is used as a benchmark rather than as a unique prediction, and the dependence on more general initial profiles has been studied in the accompanying work~\cite{Ning:2025ogq}. The curvature perturbation can induce density contrast $\delta_m \equiv \delta(t_m)$ at the horizon crossing time $t_{m}$ via $\delta_m = 4\frac{3(1+\omega)}{5+3\omega}\mu e^{-1}(1 - \mu e^{-1})$~\cite{Musco:2018rwt}. With the above initial profile, the threshold for PBH formation is $\mu_c \simeq 0.79579$ in the radiation-dominated (RD) era with $\omega=1/3$ as obtained in Ref.~\cite{Ning:2025yvj}, which corresponds to $\delta_c \simeq 0.552132$. We modify the code developed in Ref.~\cite{Escriva:2019nsa} and perform all of our simulations in the RD era, but the generalization to other $\omega$ should be straightforward. 

Some typical snapshots of our simulations with initial condition $r_m = 10 r_H$ ($r_H$ is the initial Hubble horizon) are shown in Fig.~\ref{fig:simulation} from top down for the sub-critical ($\mu=0.4$), near-critical ($\mu=0.8$), and super-critical ($\mu=0.9$) collapse cases. The corresponding radial profiles are shown in Fig.~\ref{fig:profiles} for the density contrast (top row) and comoving velocity normalized by the background fluid velocity (bottom row) in the sub-critical (left column), near-critical (middle column), and super-critical (right column) cases, respectively. Both quantities are expressed in terms of comoving radial coordinates. The shape of the sound shell depends on the amplitude of the perturbation $\mu$ at horizon reentry. A sub-critical or near-critical case will lead to sound waves with both overdense and underdense shells, for the fluid pressure gradient is large enough to push the fluid away. A super-critical case will only admit an underdense shell. 

\begin{figure*}
    \centering
    \includegraphics[width=\textwidth]{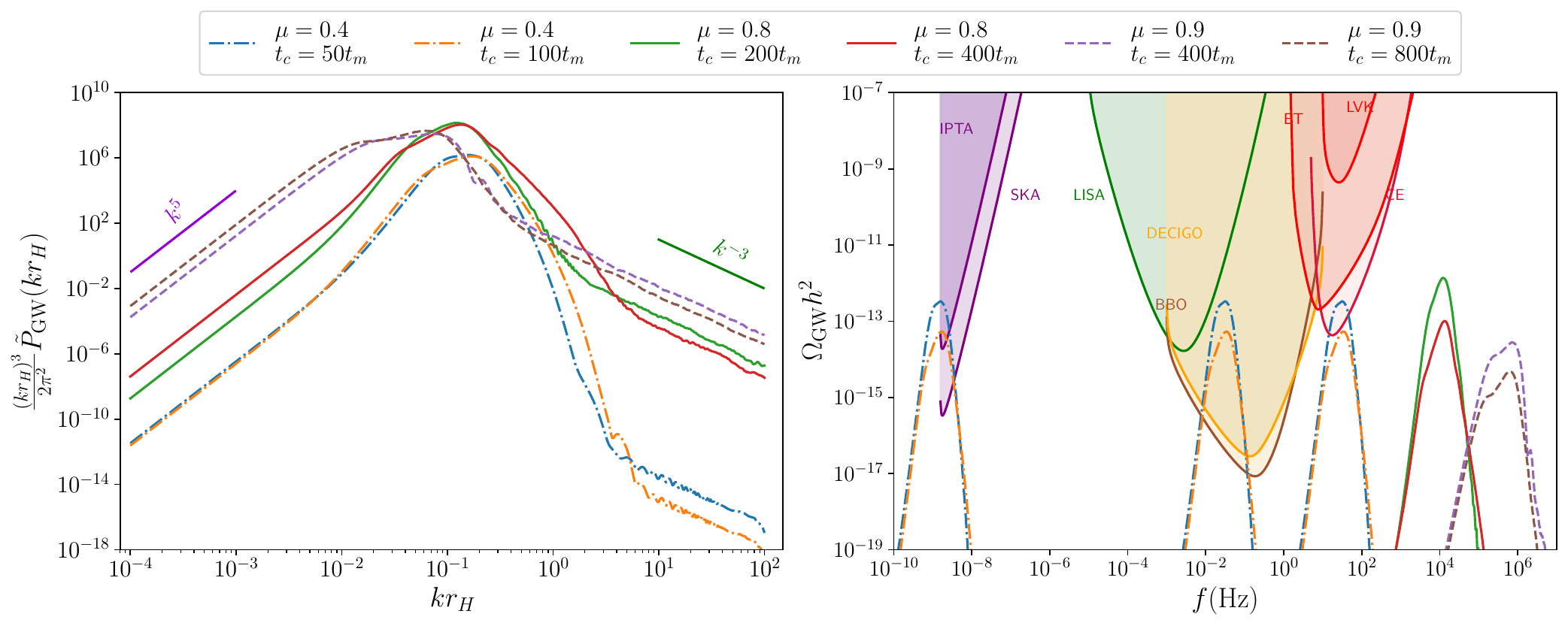}
    \caption{\textit{Left}: The dimensionless GW power spectrum. \textit{Right}: The present GW energy density spectrum. We have used~\cite{Schmitz:2020syl} to plot the sensitivity curves. In both panels, the dot-dash line denotes the sub-critical case $\mu=0.4$ with collision time $t_c = 50t_m, 100t_m$, while the solid line denotes the near-critical case $\mu=0.8$ with collision time $t_c = 200t_m, 400t_m$, and the dashed line denotes the super-critical case $\mu=0.9$ with collision time $t_c = 400t_m, 800t_m$.}
    \label{fig: OmegaGW}
\end{figure*}

\textit{\textbf{GW spectra from sound waves.}---}
Since the sound shell model~\cite{Hindmarsh:2016lnk, Hindmarsh:2019phv, Guo:2020grp}(see also Refs.~\cite{Cai:2023guc,RoperPol:2023dzg,Sharma:2023mao} for recovering the correct infrared behavior~\cite{Cai:2019cdl}) is ideally suited to our scenario, as it is specifically designed to calculate the GW signal sourced by the fluid velocity field, we closely follow this semi-analytical model to adapt to our case as detailed in the \textit{Supplementary Material} to estimate the GW spectrum. The only difference between the first-order phase transition (FOPT) case and our scenario is the input velocity power spectrum that sources the energy of GWs. For the case of a FOPT after the initial acceleration stage~\cite{Cai:2020djd}, the fluid velocity profile is self-similar~\cite{Espinosa:2010hh} driven by a constantly expanding wall until wall collision; from then on, the sound shell propagates freely with its velocity profile at the time as the input in the sound shell model. 

In our case, there is no bubble wall pushing outward the sound shell so that we can directly extract from our numerical simulations the velocity profile just a moment before the sound shell collision (to make the sound shell propagate as freely as possible) as the input in the sound shell model, and derive the fluid velocity power spectrum,
\begin{align}
    \langle v_{\vec{q}}^i v_{\vec{k}}^{j*} \rangle = \sum_{m=1}^{N}\sum_{n=1}^{N} \langle \hat{q}^i\hat{k}^j A(q)A^*(k)e^{-i\vec{q}\cdot\vec{x}^{(m)}+i\vec{k}\cdot\vec{x}^{(n)}} e^{i(\omega_q -\omega_k )t_i}\rangle,
\end{align}
where $\omega_q \equiv c_sq=q/\sqrt{3}$, $N$ is the number of sound shells at a given spatial point with their center positions labeled by $\vec{x}^{(m)}$, $t_i$ denotes the time when the sound shell starts to propagate freely and $A(q)$ encodes the information of the initial profile of density contrast and fluid velocity~\footnote{For a more accurate definition, one refers to Ref.~\cite{RoperPol:2023dzg}. Nevertheless, their correction relates to infrared behavior, and we are more concerned about the amplitude. Therefore, we ignore this correction for simplicity.},
\begin{align}
    A(q) &= \frac{1}{2}[f^{\prime}(q) + ic_sl(q)],\\
    f(q) &= \frac{4\pi}{q}\int_{r_{f}}^\infty\mathrm{d}ru(r)\mathrm{sin}(qr),\\
    l(q) &=\frac{3\pi}{q}\int_{r_f}^\infty\mathrm{d}r\delta(r)r\mathrm{sin}(qr).
\end{align}
Here $u(r)$ and $\delta(r)$ are the profiles when the sound shells start to collide, $r_f$ denotes the cutoff scale since the very high-density perturbation inside and around the PBH will not propagate outward, and we choose $r_f=20r_{\mathrm{PBH}}$ for a typical illustration. Other choices of $r_f$ only slightly change the turning-point position in the power spectrum, where the power law changes roughly from  $k^{-9}$ to $k^{-3}$) as shown in the \textit{Supplementary Material}. As for the sub-critical case, we simply take $r_f = 0$. Since we have required all sound shells to propagate freely at roughly the same time, this formalism actually assumes the monochromatic PBH formation. The ensemble average of the velocity power spectrum is taken over all sound shells centered at different positions $\vec{x}^{(m)}$ so that
\begin{align}
    \sum_{m=1}^{N}\sum_{n=1}^{N} \langle e^{-i\vec{q}\cdot\vec{x}^{(m)}+i\vec{k}\cdot\vec{x}^{(n)}} \rangle
    =(2\pi)^3\delta^3(\vec{q}-\vec{k})\frac{N}{V_c},
\end{align}
where $V_c$ is the comoving volume of space, and we can define the comoving mean separation of the sound shells in terms of $R_{*c}^{-3} \equiv N/V_c$. Finally, we arrived at
\begin{align}
    \langle v_{\vec{q}}^i v_{\vec{k}}^{j*} \rangle = \hat{q}^i\hat{q}^j|A(q)|^2(2\pi)^3\delta^3(\vec{q}-\vec{k})\frac{N}{V_c}.
\end{align}

For later convenience, we define the dimensionless velocity power spectrum as
\begin{align}\label{eq:vps}
    \mathcal{P}_v &\equiv 2\left(\frac{R_{*c}}{r_H}\right)^3\frac{q^3}{2\pi^2R_{*c}^3}|A(q)|^2 = \frac{(qr_H)^3}{\pi^2}|A(qr_H)|^2,
\end{align}
then the GW spectrum follows similarly as Ref.~\cite{Guo:2020grp},
\begin{align}
    \mathcal{P}_{\mathrm{GW}}(y, kr_H) = & 3\Gamma^2(H_sa_sR_{*c})\left(\frac{r_H}{R_{*c}}\right)^7\frac{(kr_H)^3}{2\pi^2}\nonumber\\
    &\times\Tilde{P}_{\mathrm{GW}}(kr_H)\Upsilon(y),
\end{align}
where $y$ is a time variable defined as $y=a/a_s$, $\Gamma = (\bar{\rho}+\bar{p})/\bar{\rho}\approx 4/3$, the subscript $s$ denotes the time when sound wave starts to source the GWs, roughly the time of sound shell collision, $\Upsilon$\footnote{The suppression factor $\Upsilon$ accounts for the effect of cosmological expansion on GW production, and a finite source duration only mildly modifies its value~\cite{Guo:2020grp}. However, as discussed in our second companion paper~\cite{Ning:2025yvj}, the finite lifetime of the acoustic source changes the infrared scaling to $k^3$.} is the usual suppression factor due to Hubble expansion~\cite{Guo:2020grp} that approximates to 1 at radiation dominated era and 2/3 at matter-dominated era in the long duration limit, and $\Tilde{P}_{\mathrm{GW}}$ is defined as
\begin{align}
    \Tilde{P}_{\mathrm{GW}}&(kr_H) = \frac{1}{4\pi c_s kr_H}\left(\frac{1-c_s^2}{c_s^2}\right)^2\\
    &\times\int_{z_{-}}^{z_{+}}\frac{\mathrm{d}z}{z}\frac{(z-z_{+})^2(z-z_{-})^2}{z_{+}+z_{-}-z}
    \Bar{P}_v(z)\Bar{P}_v(z_{+}+z_{-}-z) \nonumber
\end{align}
with $z = qr_H$, $z_{\pm} = \frac{kr_H}{2c_s}(1\pm c_s)$, and $\Bar{P}(z) = \frac{\pi^2\mathcal{P}_v(z)}{z^3}$. In the left panel of Fig.~\ref{fig: OmegaGW}, we show the dimensionless GW power spectrum $\frac{(kr_H)^3}{2\pi^2}\Tilde{P}_{\mathrm{GW}}(kr_H)$ at the collision time $t_c=50t_m,\,100t_m$ with a sub-critical $\mu=0.4$, $t_c=200t_m,\,400t_m$ with a near-critical $\mu=0.8$, and $t_c=400t_m,\,800t_m$ with a super-critical $\mu=0.9$, where $t_m$ is the horizon crossing time. The peak frequency and amplitude from the above analytical estimations have been verified explicitly with hydrodynamical simulations in our second accompanying paper~\cite{Ning:2025yvj}.




\textit{\textbf{Observational perspectives.}---}
As shown in the \textit{Supplementary Material}, the peak frequency in the GW spectrum is about a quarter~\cite{Ning:2025ogq} of the comoving thickness $d_c$ of the underdense shell at collisions, so that we can define a nearly constant $m\equiv d_cf_{\mathrm{peak}}\simeq\mathcal{O}(1)$. We should stress that this is not in the thin-shell approximation regime, and $d_c$ is not a free parameter, which emerges dynamically from the nonlinear evolution of the initial curvature perturbation. On the other hand, the sound shell model has found that the peak frequency is roughly the sound-shell thickness at free propagation, which has been found in our accompanying long paper~\cite{Ning:2025ogq} that $d_c$ will approach a constant proportionality to the comoving Hubble horizon $\mathcal{H}_{f}$ at reentry of initial superhorizon perturbations. Therefore, after characterizing $d_c \equiv n\mathcal{H}_{f}^{-1}$ with $n\simeq\mathcal{O}(1)$, we can estimate the present-day peak frequency $f_0=(a(t_c)/a_0)f_\mathrm{peak}=(a_f/a_0)(m/n)(H_f/2\pi)$ in terms of PBH mass $M_{\mathrm{PBH}}=4\pi\gamma M_\mathrm{Pl}^2/H_f$~\cite{Carr:1974nx, Hawking:1971ei,Sasaki:2018dmp} as
\begin{align}
f_0\simeq1.105\times10^{-2}\,&\mathrm{Hz}\left(\frac{106.75}{g_s(T_f)}\right)^\frac{1}{12}\left(\frac{g_s(T_0)}{3.94}\right)^\frac13\left(\frac{\gamma}{0.2}\right)^\frac12\nonumber\\
&\times\left(\frac{10^{20}\mathrm{g}}{M_\mathrm{PBH}}\right)^\frac12 \left(\frac{4}{n}\right)\left(\frac{m}{4}\right).
\end{align}

Therefore, for PBHs of asteroid mass $M_{\mathrm{PBH}}\simeq10^{20}\,\mathrm{g}$ within the currently open window $(10^{17}-10^{22}\,\mathrm{g})$ with maximally allowable PBH abundance $f_\mathrm{PBH}=1$, the corresponding peak frequency $f_0\simeq 1.1\times10^{-2}\,\mathrm{Hz}$ is within the LISA/Taiji/TianQin-DECIGO/BBO frequency range; while for extremely light PBHs of mass $M_{\mathrm{PBH}} < 10^{10}\,\mathrm{g}$ that are currently unconstrained, the corresponding peak frequency $f_0\simeq 1.1\times10^{3}\mathrm{Hz}$ is exactly out of the reach of LIGO-Virgo-KAGRA frequency range, that is to say, future GW detectors sensitive to higher frequencies~\cite{Aggarwal:2020olq} would, in turn, constrain these extremely light PBHs of mass $M_{\mathrm{PBH}} = 10^{-5} - 10^{9}\,\mathrm{g}$ that would emit Hawking radiation of particles beyond the current standard model of particle physics. For slightly heavier PBHs of mass $M_{\mathrm{PBH}} = 10^{10}-10^{13}\,\mathrm{g}$ constrained by Big Bang nucleosynthesis, the corresponding peak frequency $f_0\sim10-100\,\mathrm{Hz}$ is fully within the LIGO-Virgo-KAGRA frequency range. As for the sub-critical case, one can similarly obtain
\begin{align} f_0\simeq1.20\times10^{6}&\,\mathrm{Hz}\left(\frac{106.75}{g_s(T_f)}\right)^\frac{1}{12}\left(\frac{g_s(T_0)}{3.94}\right)^\frac13\nonumber\\
&\times\left(\frac{H_f}{0.05 \mathrm{Mpc}^{-1}}\right)^\frac12 \left(\frac{4}{n}\right)\left(\frac{m}{4}\right).
\end{align}

The amplitude is highly correlated to the mean separation $R_{*c}$ of the sound shell, as there is a factor $(r_H/R_{*c})^7$ in $\mathcal{P}_{\mathrm{GW}}$. The GW energy spectrum today is expressed as
\begin{align}
    h^2\Omega_{\mathrm{GW}}(f) =1.66\times 10^{-5}\left(\frac{100}{g_s(T_e)} \right)^{1/3} \mathcal{P}_{\mathrm{GW}}(a_ef/a_0),
\end{align}
with the subscript $e$ for the time of GW production. In the right panel of Fig.~\ref{fig: OmegaGW}, we show the present GW energy density spectrum with collision time $t_c=50t_m$, $t_c=100t_m$, $t_c=200t_m$, $t_c=400t_m$, and $t_c=800t_m$, respectively. The comoving mean separation of the sound shell is roughly $R_{*c} = 134r_H$, $R_{*c} = 166r_H$, $R_{*c} = 200r_H$, $R_{*c} = 280r_H$, and $R_{*c} = 380r_H$. As one can see, a larger $R_{*c}$ or a smaller $\mu$ is usually correlated to a smaller amplitude of the GW energy spectrum.

\textit{\textbf{Constraints on PBH abundance.}---}
One application of this new source of SGWBs is to constrain the abundance of PBHs, especially those that have already evaporated today (see also Ref.~\cite{Franciolini:2023osw}). Although BBN constrains the abundance of PBHs in the mass range \(10^{10}\,\mathrm{g} \lesssim M_{\rm PBH} \lesssim 10^{13}\,\mathrm{g}\), the robust constraints on PBHs with \(M_{\rm PBH}\lesssim10^{9}\,\mathrm{g}\) are still lacking. To investigate the PBH abundance, it is more convenient to use the density fraction of PBHs at the epoch of their formation,
\begin{align}
    \beta\equiv\left. \frac{\rho_{\mathrm{PBH}}}{\rho_{\mathrm{FRW}}}\right|_{a=a_{\mathrm{f}}} = \frac{n_{\mathrm{PBH}}M_{\mathrm{PBH},i}}{n_{\mathrm{H}}M_{\mathrm{H}}}=\frac{\gamma}{R_{*}^3 H_{f}^3},
\end{align}
where the subscript $f$ denotes the time when the perturbation re-enters the horizon, and $M_{\mathrm{PBH},i}$ is the initial mass of PBHs given by $M_{\mathrm{PBH},i} = \gamma M_{\mathrm{H}}|_{a_f}$. In our simulation, the collapsing efficiency $\gamma \simeq 0.355$ and $\mathcal{H}_f^{-1} = 10r_H$. Therefore, we can obtain a simple relation between $\beta$ and $R_{*c}$, $\beta = 355(r_H/R_{*c})^3$, so that $t_c=200t_m$ corresponds to $\beta = 4.4\times 10^{-5}$, $t_c=400t_m$ corresponds to $\beta = 1.6\times 10^{-5}$, and $t_c=800t_m$ corresponds to $\beta = 6.5\times 10^{-6}$. In the right panel of Fig.~\ref{fig: OmegaGW}, we have set $M_{\mathrm{PBH}}\simeq 10^8\,\mathrm{g}$ in the $\mu = 0.8$ case and $M_{\mathrm{PBH}}\simeq 10^4\,\mathrm{g}$ in the $\mu = 0.9$ case. If future observations reveal null detection for the corresponding high-frequency GWs, then the PBH abundance $\beta$ could be constrained to be smaller than $\mathcal{O}(10^{-5})$.

\textit{\textbf{Conclusions and discussions.}---}
It is of essential importance to thoroughly study all possible GW sources so as to fully understand the SGWBs in GW detectors. In this \textit{Letter}, we have proposed a hybrid numerical analysis of SGWB sourced by the collisions of sound shells from density perturbation collapse, providing an alternative probe into the evaporated PBH mass and abundance. We note here that a previous work~\cite{DeLuca:2019llr} has also studied the GW production during the non-spherical collapses of sub-critical scalar perturbations, which is overwhelmed by those SIGWs at second-order in perturbations~\cite{Ananda:2006af,Baumann:2007zm} (see Ref.~\cite{Domenech:2021ztg} for a recent review and references therein), while our work provides the first step to include the non-perturbative dynamics of the source. We found a distinctive feature of GWs sourced by super-critical PBHs and near-critical PBHs, providing a potential probe to distinguish them. The detailed comparison between our acoustic GWs and SIGWs can be found in Ref.~\cite{Ning:2025yvj} and also the \textit{Supplemental Material}.

Several aspects of this work could be further explored in future studies: (i) Our evaluation is based on monochromatic density perturbations, assuming that all sound shells are uniformly distributed and formed at roughly the same time. It is natural to consider a spatial or mass distribution, as there might be some cluster effects, which may contribute to a higher amplitude. For a broad power spectrum, the resulting distribution of sound-shell thicknesses will produce a smoother and broader peak, while its amplitude will remain controlled by the abundance of the sound shells or PBHs. In our accompanying work~\cite{Ning:2025yvj}, we have taken a first step toward relaxing the monochromatic assumption by allowing a distribution of amplitudes $\mu$ at fixed characteristic scale. We find that changing the amplitude distribution modifies the overall amplitude and ultraviolet shape, while the causal infrared behavior remains robust. A simultaneous distribution in both amplitude and characteristic size would constitute the natural next generalization; (ii) As most of our evaluation is numerical, we expect an analytical or semi-analytical profile of sound waves from density perturbation collapse, for this can give us a more general analysis of the final GW energy spectrum; (iii) Our theoretical frame can be applied to other cases where there is a sound wave structure, like PBHs formed from primordial true vacuum bubbles that later become false vacuum bubbles in the radiation era~\cite{Deng:2018cxb}; (iv) As one can see in Fig.~\ref{fig:profiles}, the sound wave cannot propagate freely when the separation of the sound shells $R_{*c}$ is too small, which should be considered separately; (v) Not only the sound shell-sound shell collisions but also the sound shell-PBH collisions will generate GWs similar to a recent numerical simulation~\cite{Yuwen:2024gcf} of bubble wall-PBH collisions.

\begin{acknowledgments}
\textit{\textbf{Acknowledgements.}---}
This work is supported by the National Key Research and Development Program of China Grants No. 2021YFC2203004, No. 2021YFA0718304, and No. 2020YFC2201501, the National Natural Science Foundation of China Grants No. 12422502, No. 12547110, No.12588101, No. 12235019, and No. 12447101, and the China Manned Space Program Grant No. CMS-CSST-2025-A01. HD was supported by Yuri Levin’s Simons Investigator Grant PG012519.
\end{acknowledgments}

\bibliography{ref}


\onecolumngrid
\newpage
\appendix

\begin{center}
{\LARGE\textbf{Supplemental Material for \\Relic gravitational waves from primordial gravitational collapses}}
\end{center}

\begin{center}
{\Large
Xiang-Xi Zeng$^{1,2}$ Zhuan Ning,$^{3,1,4}$ Zi-Yan Yuwen,$^{1,2,5}$ \\Shao-Jiang Wang,$^{1,6}$ Heling Deng,$^{7}$ and Rong-Gen Cai$^{8}$}
\end{center}

\begin{center}
{\large\textit{$^1$Institute of Theoretical Physics, Chinese Academy of Sciences (CAS), Beijing 100190, China}}
\end{center}

\begin{center}
{\large\textit{$^2$School of Physical Sciences, University of Chinese Academy of Sciences (UCAS), Beijing 100049, China}}
\end{center}

\begin{center}
{\large\textit{$^3$School of Fundamental Physics and Mathematical Sciences, Hangzhou Institute for Advanced Study (HIAS), University of Chinese Academy of Sciences (UCAS), Hangzhou 310024, China}}
\end{center}

\begin{center}
{\large\textit{$^4$University of Chinese Academy of Sciences (UCAS), Beijing 100049, China}}
\end{center}

\begin{center}
{\large\textit{$^5$Department of Physics, Stellenbosch University, Matieland 7602, South Africa}}
\end{center}

\begin{center}
{\large\textit{$^6$Asia Pacific Center for Theoretical Physics (APCTP), Pohang 37673, Korea}}
\end{center}

\begin{center}
{\large\textit{$^7$Department of Astronomy, Columbia University, New York, NY 10027, USA}}
\end{center}

\begin{center}
{\large\textit{$^8$Institute of Fundamental Physics and Quantum Technology, Ningbo University, Ningbo, 315211, China}}
\end{center}

\section{Abstract}

This Supplemental Material is prepared for the technical details of the dimensionless gravitational-wave (GW) power spectrum from the primordial gravitational collapse and its comparison to the scalar-induced GWs at second-order in perturbations. The details for implementing long-time numerical simulations of primordial gravitational collapses of density perturbations are presented in a separate accompanying paper~\cite{Ning:2025ogq} and tested under more realistic and general circumstances in a second accompanying paper~\cite{Ning:2025yvj}, where a rigorous hydrodynamic simulation extraction of acoustic GWs has perfectly matched our analytical estimation, and a more detailed comparison with SIGWs is also discussed.

\section{Appendix A: GW power spectrum from the sound shell model}
In this appendix, we closely follow the original sound shell model~\cite{Hindmarsh:2016lnk, Hindmarsh:2019phv,Guo:2020grp}. Neglecting the irrelevant scalar and vector perturbations, the metric in the FLRW Universe is defined as:
\begin{align}
    \mathrm{d}s^2 = -\mathrm{d}t^2 + a(t)^2(\delta_{ij} + h_{ij}(t, \vec{x})) \mathrm{d}\vec{x}^2,
\end{align}
where the $h_{ij}$ denotes the transverse traceless tensor perturbation. The GWs are sourced by the transverse traceless part of the energy-momentum tensor $a^2\pi^{\mathrm{TT}}$ via
\begin{align}\label{eq:eom}
    h_{ij}^{\prime\prime}(t, \vec{k}) + 2\mathcal{H}h_{ij}^{\prime}(t, \vec{k}) + k^2h_{ij}(t, \vec{k}) = 16\pi Ga^2\pi_{ij}^{\mathrm{TT}}(t, \vec{k}),
\end{align}
where $\mathcal{H}=a^{\prime}/a$, $\prime\equiv \partial/\partial \eta$ and $\eta$ is the conformal time. The GW energy density $\rho_{\mathrm{GW}}$ is defined as
\begin{align}
    \rho_{\mathrm{GW}}(t) = \frac{1}{32\pi G}\langle \dot{h}_{ij}(t, \vec{x})\dot{h}_{ij}(t, \vec{x}) \rangle.
\end{align}
Here, the dot denotes the derivative with respect to the coordinate time. Then we define the power spectrum of $\dot{h}_{ij}$ as
\begin{align}
    \langle \dot{h}_{ij}(t, \vec{k})\dot{h}_{ij}(t, \vec{q}) \rangle = (2\pi)^3 \delta^3(\vec{k} + \vec{q}) P_{\dot{h}}(t, k),
\end{align}
and the dimensionless GW energy spectrum follows
\begin{align}
    \mathcal{P}_{\mathrm{GW}}(t, k)\equiv \frac{\mathrm{d}\Omega_{\mathrm{GW}}(t) }{\mathrm{d}\ln k} =\frac{1}{\rho_c(t) }\frac{\mathrm{d}\rho_{\mathrm{GW}}(t)}{\mathrm{d}\ln k} =
    \frac{k^3}{24\pi^2 H^2}P_{\dot{h}}(t, k)
    =\frac{k^3}{24\pi^2 H^2a^2}P_{h^{\prime}}(\eta, k),
\end{align}
where the $\rho_c$ is the critical energy density at time $t$. To solve the equation~\eqref{eq:eom}, one can use the Green function method by introducing a Green function $G(x, x_0)$ with the following boundary conditions,
\begin{align}
    G(x\leq x_0) = 0, ~~~~~\left. \frac{\partial G(x,x_0)}{\partial x}\right|_{x=x_0^{+}} =1,
\end{align}
where $x = k\eta$ and $x_0$ is the time when the sound wave starts to propagate. Using the Green function, one can obtain
\begin{align}
    h_{ij}(t, \vec{k}) = 16\pi G\int_{x_0}^{x}\mathrm{d}x_1 G(x,x_1)\frac{a^{2}(\eta_1)\pi^{\mathrm{TT}}_{ij}(\eta_1, \vec{k})}{k^2}, 
\end{align}
and the 2-point correlation function is expressed as
\begin{align}
    \langle h^{\prime}_{ij}(\eta, \vec{k})h^{\prime}_{ij}(\eta, \vec{q}) \rangle = (16\pi G)^2\int_{x_0}^{x}\mathrm{d}x_1\int_{x_0}^{x}\mathrm{d}x_2 \frac{\partial G(x,x_1)}{\partial x}\frac{\partial G(x,x_2)}{\partial x} \frac{a^2(\eta_1)a^2(\eta_2)}{q^2}\langle \pi^{\mathrm{TT}}_{ij}(\eta_1,\vec{k})\pi^{\mathrm{TT}}_{ij}(\eta_2,\vec{q}) \rangle.
\end{align}
Therefore, one should evaluate the unequal time correlator (UETC) of $\pi_{ij}^{\mathrm{TT}}$. Due to the spatial homogeneity of the Universe, one can parameterize it as
\begin{align}
    \langle \pi^{\mathrm{TT}}_{ij}(\eta_1,\vec{k})\pi^{\mathrm{TT}}_{ij}(\eta_2,\vec{q}) \rangle = \Pi^2(k, \eta_1, \eta_2)(2\pi)^2\delta^3(\vec{k} + \vec{q}).
\end{align}
The only contributing part of $\pi_{ij}^{\mathrm{TT}}$ is
\begin{align}
    \pi_{ij}^{\mathrm{TT}} = (p + \rho)\gamma^2v^{i}v^j,
\end{align}
where $p$ is the pressure, $\rho$ is the energy density, $v^i$ is the velocity defined by $v^i=\mathrm{d}x^i/\mathrm{d}\eta$, and $\gamma$ is the Lorentz factor that can be taken as 1 in the non-relativistic case. As for the velocity field, we use $\Tilde{v}_{\vec{k}}^i$ to denote the Fourier transformation of $v^i(\vec{x})$. Neglecting the rotational component~\cite{Hindmarsh:2019phv}, the 2-point correlator can be defined as
\begin{align}
    \langle \Tilde{v}^i_{\vec{k}}(\eta_1)\Tilde{v}^{j*}_{\vec{q}} (\eta_2) \rangle = (2\pi)^3\delta^3(\vec{q}-\vec{k})\hat{q}^i\hat{k}^jB(q, \eta_1, \eta_2),
\end{align}
then one can directly calculate the UETC of $\pi_{ij}^{\mathrm{TT}}$ and find
\begin{align}
    \Pi^2(k, \eta_1, \eta_2) = \bar{\omega}^2\int\frac{\mathrm{d}^3q}{(2\pi)^3}B(q, \eta_1, \eta_2)B(\Tilde{q}, \eta_1, \eta_2)\frac{q^2}{\Tilde{q}^2}(1-\mu^2)^2,
\end{align}
where $\bar{\omega} = \bar{\rho}+\bar{p}$ denotes the background's homogeneous value, $\Tilde{q} = |\vec{q}-\vec{k}| $ and $\mu = \hat{q}\cdot\hat{k}$. The velocity field can be Fourier decomposed as
\begin{align}
    v^i(\eta, \vec{x}) = \frac{1}{2}\int\frac{\mathrm{d}^3q}{(2\pi)^3}[\Tilde{v}^i_{\vec{q}}(\eta)e^{i\vec{q}\cdot\vec{x}} + \Tilde{v}^{i*}_{\vec{q}}(\eta)e^{-i\vec{q}\cdot\vec{x}}],
\end{align}
where $\vec{x}$ and $\vec{q}$ are both comoving quantities. If the velocity field propagates freely as sound waves after some initial time $\eta_i$, it will also admit the plane wave solutions
\begin{align}
    v^i(\eta>\eta_i, \vec{x}) = \int\frac{\mathrm{d}^3q}{(2\pi)^3}[v^i_{\vec{q}}e^{-i\omega \eta+i\vec{q}\cdot\vec{x}} + v^{i*}_{\vec{q}}e^{i\omega\eta-i\vec{q}\cdot\vec{x}}],
\end{align}
where $\omega = c_sq$, $c_s=1/\sqrt{3}$ is the sound speed, and for the longitudinal plane wave, we have $v_{\vec{q}}^i = v_{\vec{q}}\hat{q}^i$. Note that $v^i_{\vec{q}}$ is the amplitude of the plane wave solution, which is different from $\Tilde{v}_{\vec{q}}^i$. Then one can express the $B(q, \eta_1, \eta_2)$ in terms of the 2-point function of $v_{\vec{q}}$ in the stationary case~\cite{Hindmarsh:2019phv, RoperPol:2023dzg},
\begin{align}\label{eq:defB}
    B(q, \eta_1, \eta_2) = 2P_{v}(q)\cos[\omega(\eta_1-\eta_2)],
\end{align}
where $\langle v_{\vec{q}}v_{\vec{k}}^{*} \rangle = P_{v}(q)(2\pi)^3\delta^3(\vec{k}-\vec{q})$. After defining the energy fluctuation
\begin{align}
    \lambda(x) = \frac{\rho(x)-\bar{\rho}}{\bar{\omega}},
\end{align}
the equations for sound waves follow
\begin{align}
    \Tilde{\lambda}_{\vec{q}}^{\prime} + iq^j\Tilde{v}^j_{\vec{q}} = 0,\\
    \Tilde{v}_j^{\prime} + c_siq^j\Tilde{\lambda}_{\vec{q}}=0.
\end{align}
Using these equations, one can find at some initial time $\eta_i$, the plane wave amplitude is~\cite{Hindmarsh:2019phv}
\begin{align}
    v_{\vec{q}} = \frac{1}{2}\left( \hat{q}^i\Tilde{v}^i_{\vec{q}}(\eta_i) - c_s\Tilde{\lambda}_{\vec{q}}(\eta_i) \right)e^{i\omega\eta_i}.
\end{align}
Therefore, one should assign the initial condition for the $\Tilde{v}^i_{\vec{q}}(\eta_i)$ and $\Tilde{\lambda}_{\vec{q}}(\eta_i)$. In coordinate space, the velocity profile for the $n$-th sound shell created by curvature perturbation can be written as 
\begin{align}
    v_i^{(n)}(\eta, \vec{x}) = \frac{R_i^{(n)}}{R^{(n)}}u(\eta, R^{(n)}),
\end{align}
where $\vec{R}^{(n)} = \vec{x}-\vec{x}^{(n)}$, $\vec{x}^{(n)}$ is the center of the $n$-th sound shell and $u(\eta,r)$ is radial profile of the velocity field. Assuming $\bar{\omega} = \Gamma\bar{\rho}$, one can write the energy fluctuation as
\begin{align}
    \lambda^{(n)}(\eta, \vec{x}) = \frac{1}{\Gamma}\delta^{(n)}(\eta, \vec{x}) = \frac{1}{\Gamma}\delta(\eta, R^{(n)}),
\end{align}
where $\delta(\eta, R^{(n)})$ denotes the profile of density contrast. After the Fourier transformation, one can obtain
\begin{align}
    \Tilde{v}^{i(n)}_{\vec{q}}(\eta) = e^{-i\vec{q}\cdot\vec{x}^{(n)}}i\hat{q}^if^{\prime}(\eta,q),\\
    \Tilde{\lambda}^{(n)}_{\vec{q}}(\eta) = e^{-i\vec{q}\cdot\vec{x}^{(n)}}l(\eta,q),
\end{align}
and functions $f(\eta, q), l(\eta, q)$ are given by
\begin{align}
    f(\eta, q) = \frac{4\pi}{q}\int_{0}^{\infty}\mathrm{d}r u(\eta, r)\sin (qr),\\
    l(\eta, q) = \frac{4\pi}{\Gamma q}\int_{0}^{\infty}\mathrm{d}r \delta(\eta, r)r\sin(qr).
\end{align}
It is convenient to define a function $A(\eta, r) \equiv (f^{\prime}(\eta, r)+ic_sl(\eta, r))/2$, then we have
\begin{align}
    v_{\vec{q}}^{(n)} = ie^{i\omega\eta_i-i\vec{q}\cdot\vec{x}^{(n)}}A(\eta_i, q).
\end{align}
Now we can calculate the spectral density of the plane wave amplitudes, assuming all the sound shells are propagating freely at roughly the same time $\eta_i$,
\begin{align}
    \langle v_{\vec{q}}^i v_{\vec{k}}^{j*} \rangle = \sum_{m=1}^{N}\sum_{n=1}^{N} \hat{q}^i\hat{k}^j A(q)A^*(k)\langle e^{-i\vec{q}\cdot\vec{x}^{(m)}+i\vec{k}\cdot\vec{x}^{(n)}} \rangle e^{i(\omega_q -\omega_k )\eta_i},
\end{align}
where $N$ denotes the number of the sound shell, then the ensemble average gives rise to
\begin{align}
    \sum_{m=1}^{N}\sum_{n=1}^{N} \langle e^{-i\vec{q}\cdot\vec{x}^{(m)}+i\vec{k}\cdot\vec{x}^{(n)}} \rangle
    =(2\pi)^3\delta^3(\vec{q}-\vec{k})\frac{N}{V_c},
\end{align}
where $V_c$ is the comoving volume of space. Next, we can define the comoving mean separation of the sound shell in terms of $R_{*c}^{-3} \equiv N/V_c$. Finally, we arrived at
\begin{align}
    \langle v_{\vec{q}}^i v_{\vec{k}}^{j*} \rangle = \hat{q}^i\hat{q}^j|A(q)|^2(2\pi)^3\delta^3(\vec{q}-\vec{k})\frac{N}{V_c}.
\end{align}
The dimensionless velocity power spectrum is defined as\footnote{In our main text, we have defined $\mathcal{P}_v  \equiv 2\left(\frac{R_{*c}}{r_H}\right)^3\frac{q^3}{2\pi^2R_{*c}^3}|A(q)|^2$ since this is more convenient to see the behavior of $\frac{(kr_H)^3}{2\pi^2}\Tilde{P}_{\mathrm{GW}}(kr_H)$ without the influence of $R_{*c}$.}
\begin{align}
    \mathcal{P}_{v}(q) \equiv \frac{q^3}{2\pi^2}[2P_{v}(q)]
    = \frac{q^3}{\pi^2R_{*c}^3}|A(q)|^2,
\end{align}
where the factor 2 comes from \eqref{eq:defB}. After solving the Green function~\cite{Guo:2020grp}, for example, in RD dominated era~\footnote{For the general solution, one can refer to Ref.~\cite{Guo:2020grp}.} as
\begin{align}
    G(x, x_0) = \frac{x_0\sin(x-x_0)}{x},
\end{align}
one can derive the final dimensionless GW energy spectrum
\begin{align}
    \mathcal{P}_{\mathrm{GW}}(y, kr_H) = 3\Gamma^2(H_sa_sr_H)\frac{(kr_H)^3}{2\pi^2}
    \Tilde{P}_{\mathrm{GW}}(kr_H)\Upsilon(y),
\end{align}
where $y$ is a time variable defined as $y=a/a_s$, $r_H$ is some characteristic length scale, the subscript $s$ denotes the time when the sound wave starts to source the GWs, roughly the time of sound shell collision, $\Upsilon$ is the usual suppression factor due to Hubble expansion~\cite{Guo:2020grp} that approximates to 1 at radiation dominated era and 2/3 at matter-dominated era in the long duration limit, and $\Tilde{P}_{\mathrm{GW}}$ is defined as
\begin{align}
    \Tilde{P}_{\mathrm{GW}}(kr_H) = \frac{1}{4\pi c_s kr_H}\left(\frac{1-c_s^2}{c_s^2}\right)^2
    \int_{z_{-}}^{z_{+}}\frac{\mathrm{d}z}{z}\frac{(z-z_{+})^2(z-z_{-})^2}{z_{+}+z_{-}-z}
    \Bar{P}_v(z)\Bar{P}_v(z_{+}+z_{-}-z).
\end{align}
with $z = qr_H$, $z_{\pm} = \frac{kr_H}{2c_s}(1\pm c_s)$, and $\Bar{P}(z) = \frac{\pi^2\mathcal{P}_v(z)}{z^3}$.

\section{Appendix B: The dimensionless GW power spectrum with different cutoff scale $r_f$}\label{app:peakbehaviors}

\begin{figure}[h]
    \centering
    \includegraphics[width=\linewidth]{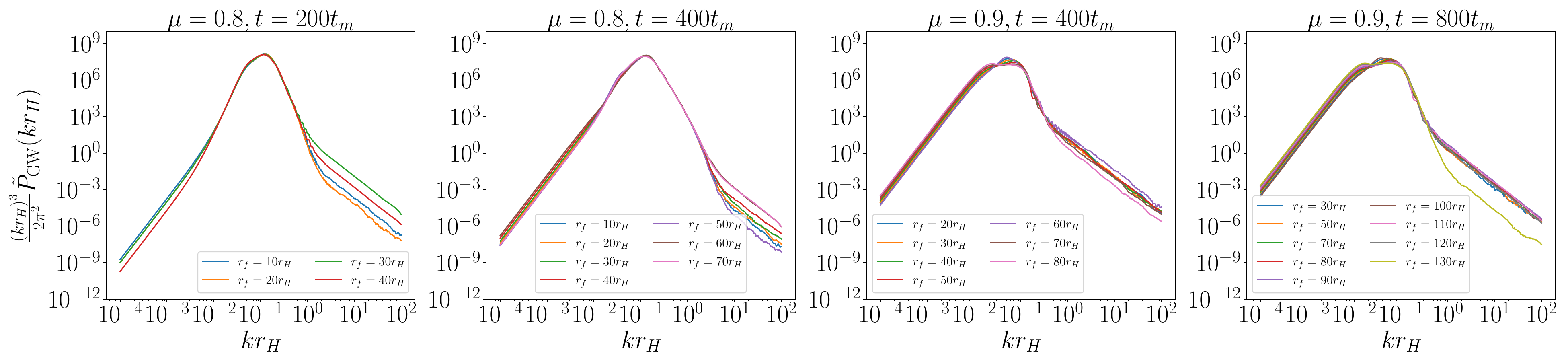}
    \caption{The dimensionless GW power spectrum with different cutoff scale $r_f$.}
    \label{fig: dimensionless_gw}
\end{figure}

As one can see in Fig.~\ref{fig: dimensionless_gw}, the peak amplitude of the dimensionless power spectrum is not sensitive to the choice of cutoff scale $r_f$, which only slightly affects the broken power law position where the power law changes from roughly $k^{-9}$ to $k^{-3}$.

\section{Appendix C: The peak behavior of the GW power spectrum}

\begin{figure}
    \centering
    \includegraphics[width=0.4\linewidth]{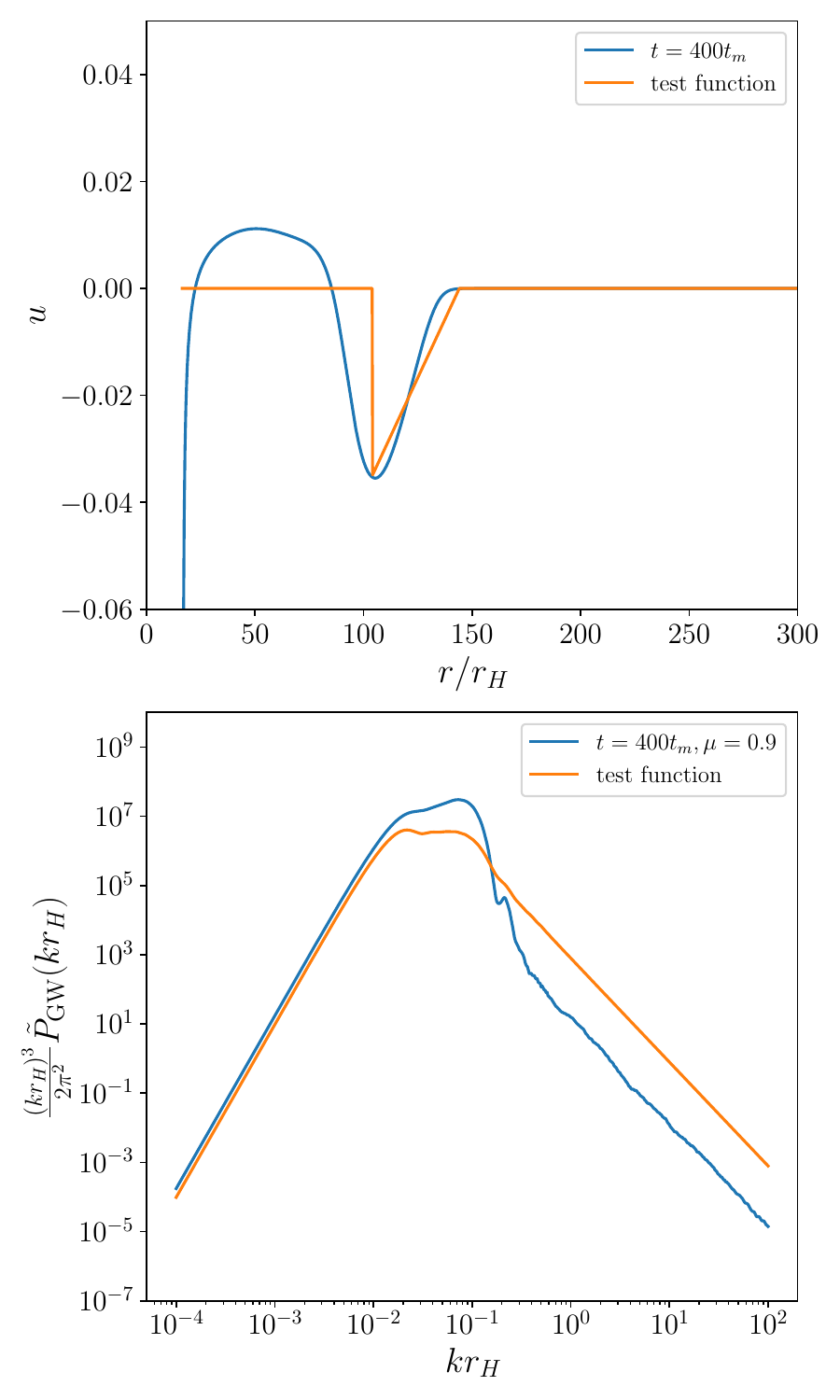}
    \hspace{0.5in}
    \includegraphics[width=0.4\linewidth]{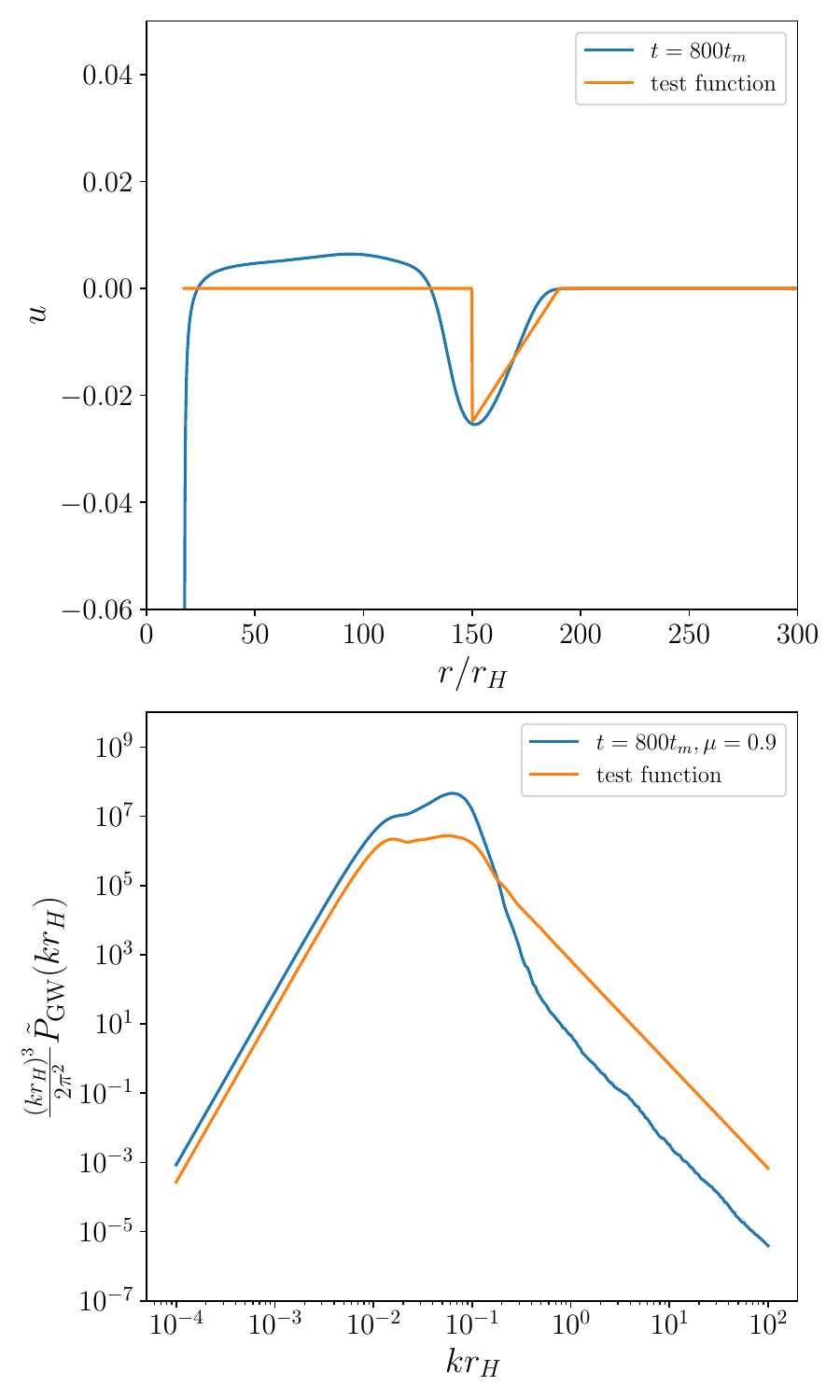}
    \caption{\textit{Left column}: The test function with $a=0.035, l=40, r_0=144$. \textit{Right column}: The test function with $a=0.025, l=40, r_0=190$. The top panels show the velocity profile, while the bottom panels show the dimensionless GW power spectrum.}
    \label{fig:test_fun}
\end{figure}

Similar to Ref.~\cite{Hindmarsh:2016lnk}, we can analyze the peak position and power law of the final GW power spectrum. Starting from the super-critical case. We use a test function,
\begin{align}
    u(r) = \frac{a}{l}(r-r_0)\Theta(r-r_0+l)\Theta(r_0-r),
\end{align}
to characterize the key features of the sound shell profile, for example, the position $r_0$ of the wavefront, the thickness $l$ (roughly at the same order of the sound-shell thickness $d_c$ when propagating freely until collision), and depth $a$ of the underdense region, where $\Theta$ is the step function and the sub-dominated overdense region is neglected for simplicity. We will use the approximated relation~\footnote{In our numerical calculation, we will directly use the initial profile of $\delta(r)$ from numerical simulations, and we will only use this approximated relation in the analytical analysis in the Supplementary Material.} $\delta(r)=4c_su(r)$ found in Ref.~\cite{Deng:2021edw}, which is consistent with our numerical results. Here are two examples shown in Fig.~\ref{fig:test_fun}.

For the velocity power spectrum $\mathcal{P}_v$ of a form $(qr_H)^n$ over a range of wave numbers, the GW power spectrum $\mathcal{P}_{\mathrm{GW}}(kr_H)$ becomes $(kr_H)^{2n-1}$. After some algebra, one can obtain
\begin{align}
    f^{\prime}(q) = \frac{4\pi a}{lq^4}\bigg{(}qr_0\cos(qr_0) + 
    q(3l-r_0)\cos(q(r_0-l)) -3\sin(qr_0) + (3+lq^2(r_0-l))\sin(q(r_0-l)) \bigg{)},
\end{align}
\begin{align}
    l(q) = \frac{4\sqrt{3}\pi a}{lq^4}\bigg{(} 2\cos(qr_0) +
    (lq^2(l-r_0)-2)\cos(q(r_0-l)) + qr_0\sin(qr_0) + q(2l-r_0)\sin(q(r_0-l))\bigg{)},
\end{align}
then we can calculate $\mathcal{P}_v$ at different limits,
\begin{align}
\mathcal{P}_v(qr_H)=\left\{
\begin{aligned}
    &\frac{a^2(qr_H)^3}{4}\left(\frac{(qr_H)^2}{225}(15l^3r_0-20l^2r_0^2+10lr_0^3-4l^4)^2+\frac{1}{36}(16l^2r_0-12lr^2_0-6l^3)^2\right),~~~~~ ql,qr_0\ll 1;\nonumber\\
    &a^2l^2r_0^2qr_H, ~~~~~ql\ll 1 \ll qr_0;\nonumber\\
    &4a^2(r_0-l)^2(qr_H)^{-1}, ~~~~1\ll ql, qr_0.
\end{aligned}
\right.
\end{align}
finally, we recover the power law behavior of $\mathcal{P}_{\mathrm{GW}}(kr_H)$ as
\begin{align}
\mathcal{P}_{\mathrm{GW}}(kr_H)\sim\left\{
\begin{aligned}
    &(kr_H)^5,~~~~~ kl,kr_0\ll 1,\nonumber\\
    &(kr_H)^1, ~~~~~kl\ll 1 \ll kr_0, \nonumber\\
    &(kr_H)^{-3}, ~~~~1\ll kl, kr_0.
\end{aligned}
\right.
\end{align}

It is worth noting that, in some special cases, the $(qr_H)^5$ term may dominate $\mathcal{P}_v$ when $ql,qr_0\ll 1$, which will contribute a $(kr_H)^9$ term in $\mathcal{P}_{\mathrm{GW}}(kr_H)$. Note also that, in the domain $ql\ll 1 \ll qr_0$, we cannot ignore the term like $2\pi a(ql)^2\sin(qr_0)/(q^4l)$ in $f^{\prime}(q)$ when $r_0$ is not much bigger than $l$, since this term will slightly change the power law within the range of $(qr_H)^{-1} \sim (qr_H)^1$, which will lead to a change in the power law of $\mathcal{P}_{\mathrm{GW}}(kr_H)$ within the range of $(kr_H)^{-3} \sim (kr_H)^1$. Equipped with the above results, one can easily find that the higher peak frequency is related to $l$, and the lower peak frequency is related to $r_0$. The analysis of near-critical cases and sub-critical cases is the same but much more complicated. However, it is not difficult to find that the higher peak frequency is related to the thickness of the overdense region, for this is the smallest scale in $u(r)$. Here, the thicknesses of the overdense and underdense regions are roughly the same; therefore, we can use the thickness of the underdense region to determine the peak frequency to be consistent with the super-critical cases.

\section{Appendix D: Comparison with scalar induced gravitational waves}\label{app:CSIGW}

\begin{figure}[htbp]
    \centering
    \includegraphics[width=0.3\linewidth]{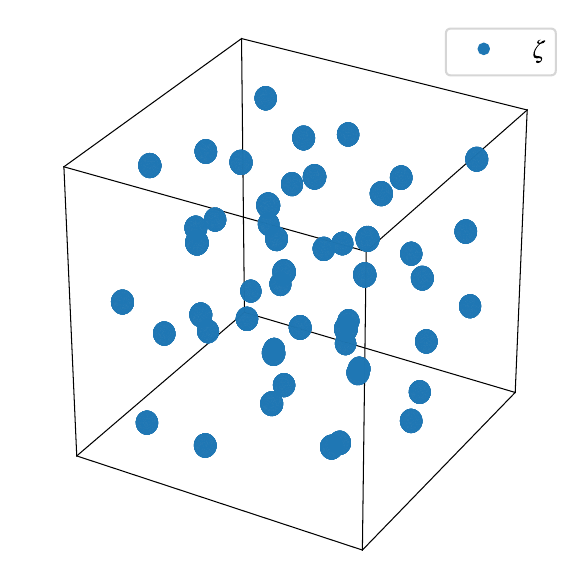}
    \caption{The initial profile of curvature perturbation $\zeta$. Blue dots denote a spherical initial Gaussian curvature perturbation with the same amplitude $\mu$.}
    \label{fig:initial}
\end{figure}

In this appendix, we will compare our acoustic GWs with the scalar-induced GWs (SIGWs) by applying the lattice simulation from the same initial profile as shown in Fig.~\ref{fig:initial}. The lattice simulation method for SIGWs is detailed in Refs.~\cite{Zeng:2025cer,Zeng:2025tno}, while the more comprehensive comparison can be found in our second accompanying paper~\cite{Ning:2025yvj}. Conventional SIGWs are obtained perturbatively: the first-order scalar metric and fluid perturbations source tensor perturbations at second order. Our acoustic GWs isolate the fluid-motion channel, but determine the fluid configuration after the large primordial peak has undergone its fully nonlinear gravitational evolution. Thus, in the perturbative regime the acoustic contribution can be viewed as the fluid part of the usual SIGW source, whereas for sufficiently large perturbations it contains nonlinear/nonperturbative fluid dynamics that are absent from the conventional second-order calculation.

To consistently compare these two kinds of signals, we have stochastically spread $N_s = 50$ initial curvature perturbation peaks in space with a Gaussian profile,
\begin{align}
    \zeta(r) = \mu e^{-(r/r_m)^2},
\end{align}
for the same amplitude $\mu$ and $r_m = 10 r_H$ in a comoving cubic box. 

\begin{figure}[htbp]
    \centering
    \includegraphics[width=0.45\linewidth]{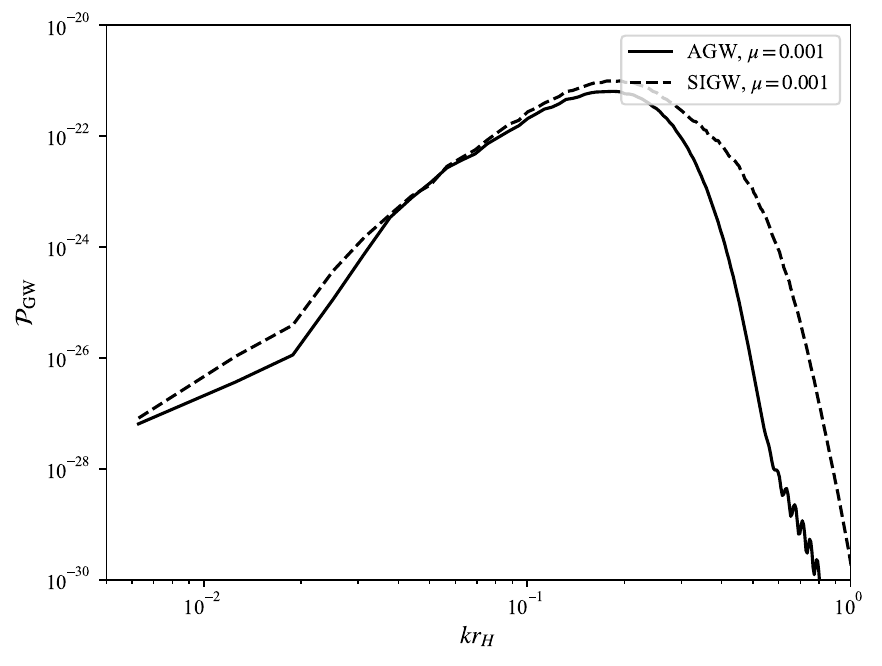}
    \qquad
    \includegraphics[width=0.45\linewidth]{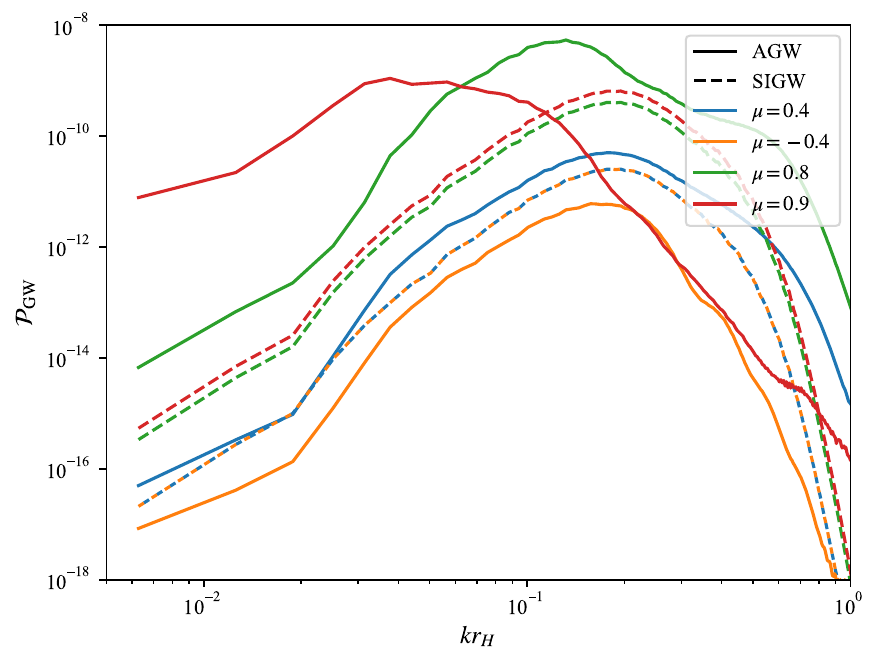}
    \caption{Comparison between the final energy spectra of Acoustic GWs (solid lines) and SIGWs (dashed lines). \textit{Left}: small perturbation with $\mu = 0.001$. \textit{Right:} Various amplitudes of the curvature perturbations: $\mu = 0.4$ (blue), $\mu = -0.4$ (orange), $\mu = 0.8$ (green), and $\mu = 0.9$ (red).}
    \label{fig:compare}
\end{figure}

For very small perturbations as shown in the left panel of Fig.~\ref{fig:compare}, the fully relativistic hydrodynamic evolution agrees with linear perturbation theory. The acoustic-GW and SIGW spectra then have similar shapes and peak locations, while the SIGW amplitude is slightly larger because SIGWs additionally contain the scalar-metric contribution. This provides an important consistency check.

For larger perturbations as shown in the right panel of Fig.~\ref{fig:compare}, the situation changes. Nonlinear gravitational evolution amplifies the fluid motion relative to its perturbative prediction. For the subcritical positive perturbations considered in our simulations, the acoustic signal can consequently exceed the perturbative SIGW result. The enhancement is strongest near the PBH-formation threshold: for our representative near-critical case, the acoustic GW amplitude exceeds the corresponding perturbative SIGW result by roughly an order of magnitude. At the same time, nonlinear broadening of the sound shell shifts the acoustic peak to a lower frequency. In the supercritical case, the signal remains enhanced relative to the perturbative result in our setup, although the disappearance of the overdense shell reduces the amplitude relative to the near-critical case. Negative curvature perturbations can instead suppress the acoustic contribution.



\end{document}